\documentclass[a4paper,twocolumn,11pt,accepted=2024-05-05]{quantumarticle}
\pdfoutput=1
\usepackage[utf8]{inputenc}
\usepackage[english]{babel}
\usepackage[T1]{fontenc}
\usepackage{amsmath}
\usepackage{hyperref}

\usepackage{amssymb}
\usepackage{amsthm}

\usepackage{tikz}
\usepackage{lipsum}

\newcommand{\cH}{{\cal H}}

\newcommand{\R}{\mathcal{R}}

\newcommand{\e}{{\rm e}}
\newcommand{\mean}[1]{\langle #1 \rangle}

\newcommand{\covsigma}[1]{\text{Cov}_\sigma[#1]}

\newcommand{\tr}{\mathrm{Tr}}

\newcommand{\rhoinfl}{\rho_\infty^\lambda}

\renewcommand{\em}{\it}

\newcommand{\change}
{{\marginpar{\#}}}        
\newcommand{\Ccal}{{\mathcal C}}

\newcommand{\Tr}{{\rm Tr}}

\newcommand{\bbbone}{\mathchoice {\rm 1\mskip-4mu l} {\rm 1\mskip-4mu l}
{\rm 1\mskip-4.5mu l} {\rm 1\mskip-5mu l}}
\newtheorem{thm}{Theorem}
\newtheorem{prop}{Proposition}

\newcommand{\av}[1]{\langle{#1}\rangle}

\begin{document}
	
\title{Rigorous results on approach  to thermal equilibrium, entanglement, and nonclassicality of  an  optical quantum field mode scattering from the elements of a non-equilibrium quantum reservoir}
	
\author{Stephan De Bi\`evre}
 \email{stephan.de-bievre@univ-lille.fr}
\affiliation{Univ. Lille, CNRS, Inria, UMR 8524, Laboratoire P. Painlev\'e, F-59000 Lille, France}
\orcid{0000-0002-2445-2701}
\author{Marco Merkli}
\email{merkli@mun.ca}

\affiliation{Department of Mathematics and Statistics, Memorial University of Newfoundland, St. John's, NL,  A1C 5S7, 
Canada}
	\author{Paul E. Parris}
	\affiliation{Missouri University of Science and Technology, Rolla, Missouri, 65409, USA }
 \email{parris@mst.edu}

\maketitle
\begin{abstract}
Rigorous derivations of the approach of individual elements of large isolated systems to a state of thermal equilibrium, starting from  arbitrary initial states,  are exceedingly rare. This is particularly true for quantum mechanical systems.
We demonstrate here how, through a mechanism of repeated scattering, an approach to equilibrium of this type actually occurs in a specific quantum system, one that can be viewed as a natural quantum analog of several previously studied classical models. 
In particular, we consider an optical mode passing through a reservoir composed of a large number of sequentially-encountered modes of the same frequency,  each of which it interacts with through a beam splitter. We  first analyze the dependence of the asymptotic state of this mode on the assumed stationary common initial state $\sigma$ of the reservoir modes and on the transmittance $\tau=\cos\lambda$ of the beam splitters.  This analysis allow us  to establish our main result, namely that at small $\lambda$ such a mode will, starting from an arbitrary initial system state $\rho$, approach a state of thermal equilibrium even when the reservoir modes are not themselves initially thermalized.
We show in addition that, when the initial states are pure, the asymptotic state  of the optical mode is maximally entangled with the reservoir and exhibits less nonclassicality than the state of the reservoir modes. 
 \end{abstract}
\section{Introduction}

\begin{figure*}
	\centering
\hspace*{.3cm}
	\includegraphics[height=6cm, keepaspectratio]{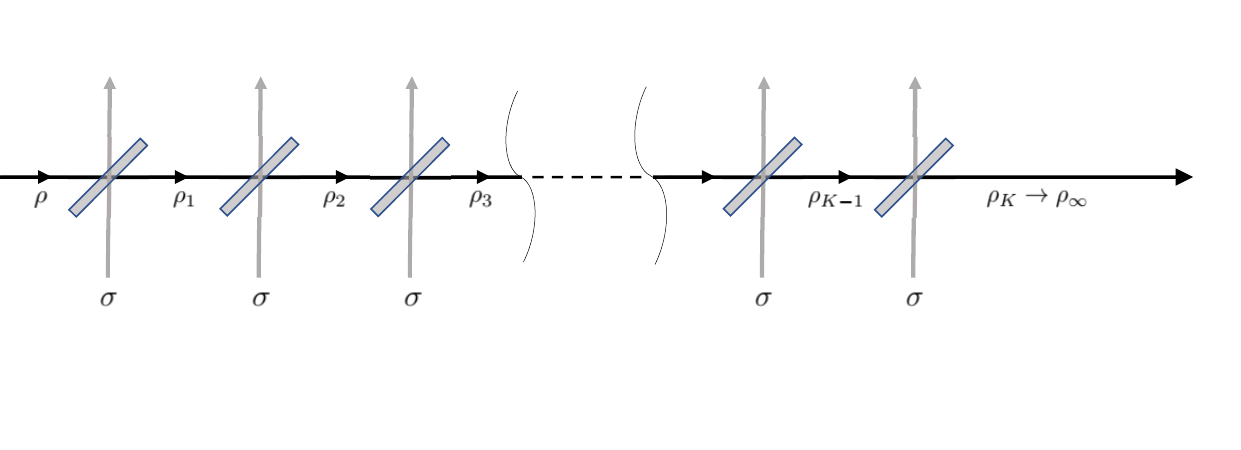}
\vspace*{-1.2cm} 
	\caption{A sequence of $K$ beam-splitters with the ``reservoir'' state $\sigma$ on each of their lower input ports and an incoming ``system'' mode in state $\rho$.}
	\label{fig:BSfigure}
\end{figure*}

 There  is continued interest (see for example~\cite{De18, Lebo93, Lebo13, vill13, Tasaki16, Mori18} and references therein) in the longstanding   problem of how  large systems, particularly quantum mechanical ones, undergo the ubiquitous process of thermalization, i.e., how it is that they are inevitably observed to approach a state of thermal equilibrium, starting from essentially arbitrary initial conditions. For large isolated quantum mechanical systems, much of this recent interest has focused on  the difficult task of verifying  the (weak or strong) Eigenstate Thermalization Hypothesis (ETH) of Deutsch  in specific systems. According to the ETH,  energy eigenstates of large systems tend, overwhelmingly, to have macroscopic properties consistent with the  thermodynamically-equilibrated states that the systems  are expected to approach. { We refer to  \cite{De18, Mori18} for  overviews of this topic and its link with the problem of thermalization. In~\cite{Mori18}, the validity of the ETH in translationally invariant lattice systems is in particular discussed. }

\color{black}
In addition to this recent progress made through a study of the validity of the ETH, it is the authors' view that valuable insight into the problem of the approach to equilibrium of large classical and quantum mechanical systems can be obtained through the identification of specific models in which an approach to equilibrium can be rigorously demonstrated. 

Recently, for example, the approach to a uniform spatial number density profile for a freely expanding classical gas has been rigorously established~\cite{DBP2017}.   Entropy growth has also been investigated~\cite{ChLeb2022} for this process, a quantum version of which has, independently, been  studied~\cite{pandey2023boltzmann} numerically.  Note, however, that in this system the absence of any interaction between the gas particles does not allow for actual thermalization, since the elements of the system do not exchange energy and momentum.

It has  on the other hand also recently been shown for several models~\cite{DBPa11, DBMePa16} that when a classical particle undergoes repeated collisions or scattering events with the local degrees of freedom of a medium through which it passes and with which it exchanges energy, the particle’s momentum and energy distribution can be driven to thermal equilibrium, even when the local degrees of freedom of the medium are not, themselves, already thermally equilibrated.

In this paper we present a fully quantum mechanical model in which a similar  mechanism of repeated scattering drives a single degree of freedom of a many-body system to thermal equilibrium.  We show that the state of the single degree of freedom converges to an equilibrium state, even though the many-body environment it is coupled to, is itself not in equilibrium. We call this dynamical process ``approach to equilibrium''. It is more intriguing than, and different from, the well-known process of ``return to equilibrium'', whereby the single degree of freedom is driven to  equilibrium when coupled to a many-body system which is itself in equilibrium.

{ The model we use belongs to a general class of models referred to as ``collision models'' or ``repeated interaction models''. They have long been used  as a versatile tool to efficiently model a variety of phenomena in equilibrium and non-equilibrium statistical mechanics, as well as in quantum information theory. An extensive overview  of the use of and physical intuition behind such collision models can be found in the recent survey paper \cite{Cicc22} and references therein. The mathematical formalism to analyse such models was developed in \cite{BJM14, BJM2006, BJM2008, BJM2010}.}

In our model, a single optical field mode (the ``system'' mode)  couples to a large reservoir of $K$ independent 
optical field modes of the same frequency through a sequence of $K$ identical beam splitters, each having a transmittance $\tau = \cos\lambda$, where the parameter $\lambda$ can be viewed as a ``coupling constant'' and where it is understood that all field modes in the system are treated quantum mechanically. (Fig.~\ref{fig:BSfigure} indicates the geometric layout, with a  horizontally propagating system mode interacting with vertically propagating reservoir modes.) 

The system mode is assumed to be in an arbitrary initial state described by a density matrix $\rho$ when it encounters the first beam splitter, at a moment when each element of the reservoir is in the same stationary  initial state $\sigma$. { The reservoir is therefore initially in a product state.} In what follows, we establish under rather mild conditions on the common initial reservoir mode state $\sigma$, that the reduced system state $\rho_K$ that emerges from the $K$-th beam splitter asymptotically converges for large $K$ to a unique limiting state 
\begin{equation}
\label{rhoinfty1}
\rhoinfl=\lim_{K\to \infty} \rho_K
\end{equation}
that generally depends on the coupling constant $\lambda$, but not on the initial state $\rho$ of the system mode. 

We then further establish that for weak coupling (i.e. small $\lambda$, corresponding to nearly perfect transmittance), the state $\rhoinfl$ that the system mode  asymptotically approaches is precisely that ``equipartitioned'' state of thermal equilibrium having the same mean energy as each of the identically-prepared reservoir modes through which it has passed.  

We note that the dynamical process we consider is unitary and hence  purely deterministic. Indeed, the only probabilistic element occurring in the model comes from that inherent in any quantum mechanical treatment.  That is to say, no {\it Stosszahlansatz} is required for demonstrating this {  instance of} approach to equilibrium.  We further stress that, since the reservoir considered in this work is not in a thermal equilibrium state, the phenomenon highlighted here  is not one of ``return to equilibrium'', which is better understood and has been much more extensively studied, including for  collision models~\cite{Cicc22, Scar02},  and more generally through the use of Lindblad equations, in the weak coupling limit.  There are also extensions to the non-equilibrium situation, where an open system is in contact with several reservoirs at different temperatures \cite{JP1, MMS}.  The continuous time limit of a non-equilibrium collision model was analyzed in \cite{KP09}, for arbitrary coupling strength.

Return to equilibrium is intuitively understood as a stability result.  The reservoir is supposed to be initially in equilibrium, while the probe degree of freedom, assumed to be weakly coupled to the reservoir,  is not, so that the state of the full system can be viewed as a small perturbation of the global equilibrium state. At long times, the full coupled system then ``returns'' to equilibrium.  Return to equilibrium occurs in the model we consider here as well. In fact, it is not limited to small coupling, but holds at all coupling strengths, as we will see below.  Moreover, we do not resort to the continuous time limit of the repeated scattering model we consider. Rather, we analyze  the discrete time collision model directly. To the best of our knowledge, the more elusive and difficult phenomenon of approach to equilibrium demonstrated here has not been previously shown to occur in collision models.
\color{black}

 Analysis of the asymptotic state for more general (e.g., non-stationary) initial reservoir states shows, moreover, that when $\rho$ and $\sigma$ are both pure states, the asymptotic system state 
$\rhoinfl$ that emerges  
is maximally entangled with the reservoir and  has less  nonclassicality than the original reservoir states.

The rest of the paper is organised as follows. In Section~\ref{s:model} the model under study is fully described.
In Section~\ref{s:convergence} we prove the existence and uniqueness of the limit implied by Eq.~\eqref{rhoinfty1} and establish properties of the asymptotic state $\rho_\infty^\lambda$. In Section~\ref{sec:nonGauss} we explore properties of the limiting state for arbitrary values of the coupling constant and show that it is Gaussian (in a sense to be defined) if the initial reservoir state $\sigma$ is itself Gaussian and provide examples when it is not. In Section~\ref{s:smallcoupling}, we study the leading order behaviour in $\lambda$ of the asymptotic state $\rho^\lambda_\infty$ for general $\sigma$, and show it is always Gaussian if $\sigma$ is, in a certain sense, centered.   We  then use the above results to establish  approach to equilibrium for the state of the system mode.
In Section~\ref{s:entnonclas} we analyse two  typical quantum mechanical properties of the asymptotic state: its entanglement to the reservoir and its nonclassicality.

\section{The Model}\label{s:model}

As described above, we consider a single mode of an optical field, with annihilation and creation operators $a, a^\dagger$, that we shall refer to as the $a$-mode or the system mode. This mode, starting from an initial input state $\rho$,  enters sequentially the ``horizontal''  input ports of a long sequence of $K$ beam splitters. (See Fig.~\ref{fig:BSfigure}) The mode at the ``vertical'' input port of the $k$-th beam splitter is characterized by annihilation and creation operators $b_k, b^\dagger_k$. 
To begin our analysis we simply assume that the reservoir mode associated with each beam splitter is initially in the same (not  necessarily stationary) state $\sigma$.    
In what follows, a state $\sigma$ is said to be  {\em stationary} if $[\sigma, b^\dagger_k b_k]=0$, and to be {\em centered} if $\langle b_k\rangle_\sigma=\langle b^\dag_k\rangle_\sigma=0$.
 The operation of the $k$-th beam splitter on the $a$-mode is determined by the unitary scattering matrix~\cite{hara13} 
\begin{equation}
S_k=\e^{\lambda (a^\dagger b_k - a b_k^\dagger)},\qquad  \mbox{where\quad  $0\leq \lambda\leq \tfrac{\pi}{2}$}
\end{equation}
and where, e.g., we  will often simply write $ab$ instead of $a\otimes b$ for the 
 tensor product of operators from different factor spaces. It follows that\footnote{One readily finds the differential equation $\frac{d^2}{d\lambda^2} S^\dag_k aS_k = -S^\dag_kaS_k$.}
\begin{equation}\label{eq:BS1}
S^\dagger_k aS_k = a \cos\lambda + b_k \sin\lambda.
\end{equation}
The state $\rho_k$ of the $a$-mode at the output of the $k$-th beam splitter can be computed recursively through the relation
\begin{equation}\label{eq:rhok}
\rho_k=\Tr_k S_k(\rho_{k-1}\otimes \sigma)S_k^\dagger,
\end{equation}
in which $\Tr_k$ denotes the partial trace over the mode corresponding to $b_k$.  Since the beam splitters are passive elements, one has
$$
[S_k, a^\dagger a+b_k^\dagger b_k]=0,
$$
so that the total photon number, and thus the total energy of the full system, is preserved in each scattering event. Not to encumber the notation unnecessarily, we suppress the $\lambda$-dependence in the notation for both  $S_k$ and  the sequentially evolving states of the  small subsystem, although this dependence is essential in what follows. 

We think of the family of modes $b_k$, each associated with an identical Fock space $\cH_k = \cH_b$, as forming a reservoir and write $\cH_{\R} = \cH_1\otimes\dots\otimes \cH_K$ for the corresponding product Fock space. The reservoir is assumed to be initially in the $K$-fold product state $\sigma\otimes\dots\otimes\sigma$. The $a$-mode, that we think of as a small subsystem, has a corresponding Fock space $\cH_{a}$  and is initially in the state $\rho$. The initial state of the total system-reservoir complex is thus the product state $\rho\otimes\sigma\otimes\cdots\otimes\sigma$.
\color{black}

The small subsystem undergoes successive interactions with each of the modes $b_k$, and in what follows we establish properties of the asymptotic state 
\begin{equation}
\label{rhoinfty}
\rhoinfl=\lim_{K\to \infty} \rho_K
\end{equation}
that it approaches as the number of beam splitters $K$ tends to infinity. This corresponds to a natural thermodynamic limit for this open system. We will see that the asymptotic state $\rho_\infty^\lambda$ depends on $\sigma$ and possibly on $\lambda$, but not on the initial state $\rho$ of the $a$-mode. 

Since the reservoir modes all start off in the same initial state $\sigma$, it is clear from Eq.~\eqref{eq:rhok} that the dynamics of the $a$-mode is given by  iteration of the following $k$-independent quantum channel:
\begin{equation}\label{eq:defL}
L(\rho) = {\rm Tr}_b\big( S (\rho\otimes \sigma) S^\dagger\big), \quad S = \e^{\lambda (a^\dagger b - a b^\dagger)},
\end{equation}
where the partial trace is taken over the single $b$-mode degree of freedom. 
After the $a$-mode has passed through $k$ beam splitters, its state $\rho_k$ is thus
\begin{equation}
\rho_k=L^k(\rho).
\end{equation}
The scattering operator $S$ is both unitary and Gaussian, where by the latter we mean that it is an exponential function of a sum of at most bi-linear products of annihilation and creation operators from any of the factor spaces.  As a result, it is convenient to characterize the states $\rho_k$ of the system mode and the initial states $\sigma$ of the reservoir by their characteristic functions. To this end we introduce the displacement operator 
\begin{equation}\label{eq:displ} 
D_{b}(z)=\exp(z b^\dagger-z^*b),\qquad z\in\mathbf C,
\end{equation}
for a mode associated with the operators $b,b^\dagger$. The characteristic function of a density matrix $\rho=\rho_b$ on $\cH_{b}$ is then defined by
\begin{equation}
\chi_{\rho}(z)=\Tr_b(\rho D_{b}(z)):=\av{D_{b}(z)}_{\rho}.
\label{charfunct}
\end{equation}
We will apply these definitions both to the system mode and to the reservoir modes and we denote the system ($a$-mode) displacement operator  by $D_a(z)$.

We will also refer to a single-mode state $\sigma$ as Gaussian~\cite{WeedbrookRMP} if its characteristic function is a Gaussian function of $z$, i.e., if
\begin{equation}
	\label{Gauss}
	\chi_\sigma(z)= {\rm Tr}_b ( \sigma D_b(z) ) =e^{G(z)}, 
\end{equation}
where
\begin{equation}
\label{exponent}
G(z)=\tfrac14 Z^T \Omega^TA\Omega Z-\Delta^T\Omega Z.
\end{equation}
Here, 
$$
Z=\begin{pmatrix} z\\  z^*\end{pmatrix},\quad \Omega=
\begin{pmatrix}
0&1\\
-1 &0
\end{pmatrix}
$$
and $A$ and $\Delta$ are, respectively, the {\it covariance matrix} and the {\it displacement vector} of $b,b^\dagger$, defined by
\begin{eqnarray}
\label{eq:Adef}
A&=&2\begin{pmatrix}\covsigma{b,b}&\covsigma{b,b^\dag}\\\covsigma{b^\dag,b}&\covsigma{b^\dag,b^\dag}
	\end{pmatrix}, \  \textrm{and}\quad \\
\Delta&=&\begin{pmatrix}\mean{b}_\sigma\\\mean{b^\dagger}_\sigma\end{pmatrix}.
\end{eqnarray}
In this last expression, 
\begin{equation}
\label{def:cov}
\covsigma{x,y} =\tfrac12\mean{xy+yx}_\sigma-\mean{x}_\sigma\mean{y}_\sigma. 
\end{equation}
Clearly, for a centered state $\Delta = 0.$

A particular case of a stationary Gausssian state is the  {\em thermal state at inverse temperature $\beta>0$}, given by the density matrix
\begin{equation}
\label{equilsigma}
\sigma_\beta:=Z_\beta^{-1}\exp(-\beta b^\dagger b),
\end{equation}
where $Z_\beta={\rm Tr}\, e^{-\beta b^\dag b}=(1-e^{-\beta})^{-1}$ is the partition function. The associated characteristic function is 
\begin{equation*}
	\av{D_b(z)}_{\sigma_\beta} =\exp \big[-\tfrac12 |z|^2\coth(\beta/2)\big].
\end{equation*}
The covariance matrix and displacement vector associated with the thermal state $\sigma_\beta$ are
\begin{equation}
\label{thermalgauss}
A_\beta = \begin{pmatrix}
   0 & 2 \overline n_\beta+1\\  
   2\overline n_\beta+1  & 0
\end{pmatrix}
\quad\mbox{and \quad $\Delta_\beta=0$,}
\end{equation}
where 
\begin{equation*}
\overline n_\beta = \av{b^\dag b}_{\sigma_\beta} = \frac{1}{e^{\beta}-1}
\end{equation*} 
is the average photon occupation number in $\sigma_\beta$.


\section{The asymptotic state and its properties}\label{s:mainresults}

\subsection{Convergence to and 
expressions for the asymptotic state $\rho_\infty^\lambda$}\label{s:convergence}
The goal of this subsection is to establish the convergence implied by Eq.~\eqref{rhoinfty} and to study some general features of the asymptotic state $\rho^\lambda_\infty$. More precisely, we will show that for {\color{black} (almost) arbitrary initial reservoir states $\sigma$ and} $0<\lambda\leq \frac{\pi}{2}$, the characteristic function after $K$ interactions has a limit as $K\rightarrow\infty$ given by the relation
\begin{eqnarray}
	\label{11}
	\chi_\infty^{\lambda}(z)&:=&\lim_{K\rightarrow\infty} \av{L^K(\rho) D_a(z)}_\rho\nonumber\\
 &=& \prod_{k=0}^{\infty} \av{D_{b}(\sin\lambda\,  [\cos\lambda]^k z)}_\sigma,
\end{eqnarray}
and that there exists an asymptotic state, i.e., an $a$-mode density matrix $\rho_\infty^\lambda$, with a characteristic function 
$\chi_\infty^\lambda(z) =  \av{D_a(z)}_{\rhoinfl}
$ given by the right hand side of this last expression.

Notice that, once established, Eq.~\eqref{11} 
implies that the asymptotic state $\rho_\infty^\lambda$ does not depend on the initial state $\rho$, {\color{black} whether the reservoir state $\sigma$ is stationary or not. Thus, all} memory of the initial system state is lost in the repeated scattering process. Equation~\eqref{11} therefore establishes that the quantum channel $L$ defined in Eq.~\eqref{eq:defL} admits a unique stationary state, asymptotically attained by the system mode after it has interacted  with many reservoir  modes.

Generally, however, the asymptotic state does depend on $\lambda$ and on $\sigma$. Indeed,
Eq.~\eqref{11} implies that (where $\partial_z=\tfrac12 (\partial_x-i\partial_y)$, $\partial_{z^*}=\tfrac12 (\partial_x+i\partial_y)$)
\begin{eqnarray}
\langle a\rangle_\infty^\lambda\,&:=&\Tr(a\rho_\infty^\lambda)\nonumber\\
&=&-\partial_ {z^*}\chi_\infty^\lambda(0)=\frac{\sin\lambda}{1-\cos\lambda}\langle b\rangle_\sigma,\label{eq:ab}\\
\langle a^\dagger\rangle_\infty^\lambda&:=&\Tr(a^\dagger\rho_\infty^\lambda)\nonumber\\
&=&\quad\partial_{ z}\chi_\infty^\lambda(0)=\frac{\sin\lambda}{1-\cos\lambda}\langle b^\dagger\rangle_\sigma.\label{eq:adagbdag}
\end{eqnarray}
It follows that the mean displacement
of the asymptotic state is, up to a $\lambda$ dependent factor, equal to that of  each of the reservoir modes.
Moreover, it follows from Eq.~\eqref{11} and a straightforward computation that 
\begin{eqnarray*}
\lefteqn{-\partial_z\partial_{z^*}\chi_\infty^\lambda(0)= -\partial_z\partial_{z^*}\chi_\sigma(0)}\\
&& -\frac{\sin \lambda}{1-\cos \lambda}\partial_{z^*}\chi_\sigma(0)\frac{\sin \lambda}{1-\cos \lambda}\partial_z\chi_\sigma(0)\\
&&+\partial_{z^*}\chi_\sigma(0)\partial_{z}\chi_\sigma(0).
\end{eqnarray*}
Consequently, using Eq.~\eqref{eq:ab} and Eq.~\eqref{eq:adagbdag}, as well as
\begin{eqnarray*}
\Tr(a^\dagger a\rho_\infty^\lambda)+\frac12 &=&-\partial_z\partial_{z^*}\chi_\infty^\lambda(0),\\
\Tr(b^\dagger b\sigma)+\frac12&=&-\partial_z\partial_{z^*}\chi_\sigma(0),
\end{eqnarray*}
one finds readily that
\begin{eqnarray*}
\mathrm{Cov}_{\rho_\infty^\lambda}[a^\dagger,a]&=&\Tr \rho_\infty^\lambda(a^\dagger-\langle a^\dagger\rangle_\infty^\lambda)(a-\langle a\rangle_\infty^\lambda)+\frac12\\
&=&\Tr\sigma(b^\dagger-\langle b^\dagger\rangle_\sigma)(b-\langle b\rangle_\sigma)+\frac12\\
&=&\mathrm{Cov}_\sigma[b^\dagger,b].
\end{eqnarray*}
In other words, second order fluctuations about the mean displacement of the asymptotic system state are the same as those of the reservoir states. They do not, therefore, depend on $\lambda$. This implies, in particular, that when the initial reservoir state $\sigma$ is  centered,  the mean photon number in the asymptotic state of the $a$-mode is the same as in each of the initial reservoir modes. We deduce, for example, that if the $a$-mode  initially has more photons than each of the reservoir modes, it will lose photons on average to the reservoir, and vice versa. 
Thus, in this situation, an {\it equipartition} of photon number and energy develops as the system mode passes through the reservoir.

Technical details of the derivation of~Eq.~\eqref{rhoinfty} and~Eq.~\eqref{11} are presented in Appendix \ref{App:A}. Here we present the essential ideas of the argument in simplified form. 

In the interest of compactness, we abbreviate $s=\sin(\lambda)$ and $c=\cos(\lambda)$, and infer from
 Eq.~\eqref{eq:BS1}  that  $S^\dagger(z a^\dagger -z^* a)S=(cz) a^\dagger -(cz)^* a + (sz)b^\dagger-(sz)^*b$, so that 
\begin{eqnarray*}
S^\dagger D_a(z)S&=&\exp\Big[ (cz) a^\dagger -(cz)^* a)\Big]\\
&&\times \exp \Big[(sz)b^\dagger-(sz)^*b) \Big]\\
&=&D_a(cz)D_b(sz).
\end{eqnarray*}
Denoting traces over the single modes $a$, $b$ and the combined modes $ab$ by ${\rm Tr}_a$, ${\rm Tr}_b$ and ${\rm Tr}_{ab}$, respectively, we then obtain
\begin{eqnarray}
	\label{eq:Laction}
\lefteqn{{\rm Tr}_{a}\big( L(\rho) D_a(z)\big)}\nonumber\\
&=& {\rm Tr}_{ab}\big( (\rho\otimes\sigma) S^\dag (D_a(z)\otimes\bbbone) S\big)\nonumber\\
	&=& {\rm Tr}_a(\rho D_a(cz))\ {\rm Tr}_b(\sigma D_b(sz)).
\end{eqnarray}
After $k$ applications, the initial density matrix $\rho$ is transformed into $L^k(\rho)=L(L^{k-1}(\rho))$. Eq.~\eqref{eq:Laction} then gives 
\begin{eqnarray}
\lefteqn{\langle D_a(z)\rangle_{L^k(\rho)}}\nonumber\\
&=&     {\rm Tr}_a\big(L^k(\rho)D_a(z) \big)\nonumber\\
  &=& {\rm Tr}_{ab}\big( (L^{k-1}(\rho)\otimes\sigma) (D_a(cz)\otimes D_b(sz)) \big)\nonumber\\
 &=& \langle D_a(cz)\rangle_{L^{k-1}(\rho)} \langle D_b(sz)\rangle_\sigma.
	\label{8}
\end{eqnarray}
Iterating this formula yields for $K\ge1$,
\begin{equation}
	\label{10}
	\langle D_a(z)\rangle_{L^K(\rho)} = \av{D_a(c^K z)}_\rho\ \prod_{k=0}^{K-1} \av{D_{b}(sc^k z)}_\sigma.
\end{equation}
In this last expression, under the  assumptions previously stated,  $c^K\rightarrow 0$ as $K\rightarrow\infty$, in which limit  Eq.~\eqref{10} reduces to Eq.~\eqref{11}. Full technical details of the derivation of the results in this section are presented in Theorem~\ref{thm:existence} of Appendix \ref{App:A}.

\subsection{Gaussian and non-Gaussian asymptotic system states
 and return to equilibrium
}\label{sec:nonGauss}
We now establish that the asymptotic system state $\rho_\infty^\lambda$ is Gaussian whenever the initial reservoir state $\sigma$ is. To this end, let $\sigma$ be a Gaussian reservoir state with covariance matrix $A$ and displacement vector $\Delta$, as defined in Eq.~\eqref{Gauss}. We then have
$$
\chi_\sigma(s c^kz) = \exp \big[\tfrac14 s^2 c^{2k}\ 
 Z^T\Omega^TA\Omega Z- sc^k\, \Delta^T\Omega  Z\big],
$$
where we have again used the abbreviations $s=\sin\lambda$ and $c=\cos\lambda$.
By Eq.~\eqref{11}, the characteristic function of the asymptotic state $\rho_\infty^\lambda$ is
\begin{eqnarray}
\chi_\infty^\lambda(z) &=& \prod_{k\ge 0}\chi_\sigma(sc^kz)\nonumber\\
&=& \exp\big[ \tfrac14 s^2\big(\sum_{k\ge 0} c^{2k} \big)   Z^T \Omega^TA\Omega Z\nonumber\\
&& \qquad - s \big(\sum_{k\ge 0} c^k \big)  \Delta^T\Omega  Z \big]\nonumber\\
&=& \exp\big[ \tfrac14 \,   Z^T \Omega^TA\Omega Z- \tfrac{s}{1-c}\,  \Delta^T\Omega  Z \big].\qquad
\label{25}
\end{eqnarray}
Equation~\eqref{25} thus establishes the following: 
\begin{prop}
\label{lem1}
If $\sigma$ is Gaussian with covariance matrix $A$ and displacement vector $\Delta$, then the asymptotic state $\rho_\infty^\lambda$ is also Gaussian, with covariance matrix $A_\infty^\lambda$ and displacement vector $\Delta_\infty^\lambda$ given by
$$
A_\infty^\lambda=A\quad \mbox{and}\quad \Delta_\infty^\lambda=\frac{\sin \lambda}{1-\cos\lambda}\Delta.
$$
\end{prop}

\noindent
Note that the Proposition holds for arbitrary (i.e., Gaussian or non-Gaussian) initial system states $\rho$.
Thus, although a non-vanishing interaction is crucial for driving the system to an asymptotic state, the covariance matrix of the latter ends up being independent of the strength of that interaction.  

As an important special case of this last Proposition note also that, 
when $\Delta=0$, so that $\sigma$ is a centered Gaussian, then $\rho_\infty^\lambda=\sigma$. This then leads to 
\begin{prop}
\label{RTE}
When a system mode in an arbitrary initial state $\rho$ passes through a reservoir, the modes of which are all in the same thermal state  $\sigma = \sigma_\beta$ at inverse temperature $\beta$ then, independent of the interaction strength $\lambda$, the system state of the $a$-mode  will  converge to  a thermal state at the same inverse temperature, i.e.,
$$
\rho_\infty^\lambda=\sigma_\beta .
$$
\end{prop}

\noindent
Thus, this fully quantum mechanical model exhibits a ``return to equilibrium'', in which the small system mode is driven to the same thermal equilibrium shared by the already equilibrated reservoir.
Similar return to equilibrium processes are familiar from  the open quantum systems literature, where one often considers the dynamics to be generated by a Hamiltonian $H=H_S+H_R+\lambda V$, consisting of a system term, a reservoir term, and an interaction term with coupling constant $\lambda$. Return to equilibrium for such systems, where the system has finitely many levels, the reservoir is thermodynamically large, and the coupling suitably small, was proven in \cite{JP96, BFS, FM, MM-AOP, MM-Quantum} (see also references therein).  The setup considered in the current manuscript is somewhat different. The dynamics of the $a$ mode (the system), given by Eq.~\eqref{eq:rhok}, is that of a {\it repeated scattering process}, as the system is in contact with fresh reservoir elements sequentially in time (each element being a $b$ mode). 
The main idea of return to equilibrium remains nevertheless the same:  a unique system mode being out of equilibrium constitutes only a small perturbation to the global equilibrium of the joint system-reservoir universe.  
Our analysis above shows that we do indeed have return to equilibrium {\em for all values of the coupling constant $\lambda$}.

To end this subsection, we consider a final example in which neither the initial reservoir state $\sigma$ nor the asymptotic state $\rho_\infty^\lambda$ is Gaussian. In particular, we consider the   case in which each of the reservoir modes is in a Fock state $\sigma=|n\rangle\langle n|$ where $|n\rangle =\frac{1}{\sqrt{n!}} (a^\dagger)^n|0\rangle$, and $n\ge 0$. From Eq.~\eqref{11} one finds for this case that
\begin{equation}
	\chi_\infty^{\lambda}(z)=\Tr (\rho_\infty^\lambda D_a(z)) = \e^{-\frac12 |z|^2} \prod_{k=0}^\infty p_n\big(|s c^k z|^2\big),
	\label{13.1}
\end{equation}
in which $p_n(x) = \sum_{j=0}^n {n\choose j}\frac{1}{j!} (-x)^j$ is the $n$th Laguerre polynomial~\cite{hara13}, which is known to be the characteristic function of the state $|n\rangle$.  It is straightforward to show that the state associated with \eqref{13.1} is then not Gaussian except when $n=0$. 
Indeed, expanding\footnote{We have $\ln[\chi_\infty^\lambda(t)] = -\frac12 t^2+\sum_{k\ge 0}\ln [p_n(s^2c^{2k}t^2)]$. By expanding $p_n(\epsilon)=1-n\epsilon+\frac12 n(n-1)\epsilon^2 +O(\epsilon^3)$ for small $\epsilon$, and then expanding the logarithm, one readily obtains the above mentioned expression.} the logarithm $\ln[\chi_\infty^\lambda(t)]$ for small $t\in\mathbf R$, it is straightforward to establish that the Taylor coefficient of $t^4$ in that expansion is 
$-\frac{n}{2}s^2$, which implies that the asymptotic state $\rho_\infty^\lambda$ is not Gaussian for $n>0$.
 
 The system mode in this case is thus not  driven to a Gaussian equilibrium state by the non-equilibrium reservoir. Nevertheless, from the previous section it follows that the mean number of photons in the asymptotic state is the same as the actual initial number $n$ of photons in each of the reservoir modes. In this sense, a form of equipartition still takes place.
For the particular case $n=1$, we explicitly obtain
\begin{equation*}
	\chi_\infty^{\lambda}(z)=\Tr (\rho_\infty^\lambda D_a(z)) 	 = \e^{-\frac12 |z|^2}  (s^2|z|^2; c^2)_\infty,
\end{equation*}
where 
$$
(a;q)_\infty \equiv \prod_{k=0}^\infty (1-aq^k)
$$
is the {$q$-Pochhammer symbol}, which defines a function of $q$ analytic in $|q|<1$. In our case, $q=c^2=\cos^2(\lambda)$.

In the next subsection, we show that even when the reservoir states $\sigma$ are not Gaussian, as long as they are centered, the asymptotic system state $\rho_\infty^\lambda$ will itself approach a Gaussian state as the coupling strength $\lambda$ goes to zero. This result will allow us to   establish that at weak coupling the system exhibits a true ``approach to equilibrium'', provided the reservoir modes are in a stationary state $\sigma$.

\subsection{The asymptotic system state at weak  coupling and approach to equilibrium
}\label{s:smallcoupling}
We now investigate the asymptotic system states $\rhoinfl$ that arise  at small values of the coupling strength $\lambda$. We first consider the situation in which the initial states $\sigma$ of the reservoir modes are  centered ($\Delta =0$), but are not necessarily Gaussian. Our main finding for this case is that the dominant term of the asymptotic state, 
\begin{equation}
\label{eqrho0}
\rho_\infty^0=\lim_{\lambda\rightarrow 0}\rho_\infty^\lambda,
\end{equation}
is a {\it Gaussian state} having zero displacement and a covariance matrix {\it equal} to that of the initial states $\sigma$ of the reservoir, even when the latter states are not, themselves, Gaussian.  
This result is proven in Appendix~\ref{app:approach}, see Theorem~\ref{thm:gaussianlim}. We sketch the main argument of the proof here.  For that purpose,  we introduce the notation
\begin{equation}
\label{varphidef}
	\varphi(z) = zb^\dag -z^*b,
\end{equation}
and then compute
\begin{eqnarray*}
	D_b(z) &=& \e^{\varphi(z)},\\
 \av{e^{\varphi(z)}}_\sigma &=& 1 +\av{\varphi(z)}_\sigma+\tfrac12 \av{\varphi(z)^2}_\sigma + O(|z|^3).
\end{eqnarray*}
We show in Appendix \ref{app:approach} (see Proposition \ref{thm2}) that 
\begin{eqnarray}\label{eq:lambdadevelop}
\lefteqn{\chi_\infty^{\lambda}(z)=\tr\big(\rho_\infty^\lambda D_a(z)\big)}\\
  &=& \exp\Big[  2\lambda^{-1}\av{\varphi( z)}_\sigma +\tfrac12 {\rm Var}_\sigma\big(\varphi(z)\big) +O(\lambda) \Big],
\nonumber
	\end{eqnarray}
 where ${\rm Var}_\sigma(X)=\av{X^2}_\sigma-\av{X}_\sigma ^2$. 
 For  centered reservoir states $\sigma$, the displacement $\Delta$ vanishes, and thus so does the mean value 
 $\av{\varphi( z)}_\sigma$. 
 For  centered  reservoir states, therefore,
\begin{eqnarray}
\label{eq:lambda0limit}
	\chi_\infty^{0}(z)&=&\tr\big(\rho_\infty^0 D_a(z)\big)\\
 &:=&\lim_{\lambda\rightarrow 0} \tr\big(\rho_\infty^\lambda D_a(z)\big)\nonumber\\
 &=& \exp\Big[  \tfrac12 {\rm Var}_\sigma\big(\varphi(z)\big) \Big].
 \nonumber
\end{eqnarray}
Furthermore, one directly verifies that 
\begin{eqnarray}
\lefteqn{{\rm Var}_\sigma\big(\varphi(z)\big)=\covsigma{\varphi(z),\varphi(z)}}\\
	&=&
	\begin{pmatrix} z & z^*\end{pmatrix}
	\begin{pmatrix}
		\covsigma{b^\dagger, b^\dagger}&-\covsigma{b^\dagger, b}\\
		-\covsigma{b,b^\dagger}&\covsigma{b,b}
	\end{pmatrix}
	\begin{pmatrix}
		z\\z^*
	\end{pmatrix},
 \nonumber
 \label{24.1}
\end{eqnarray}
where $\covsigma{x,y}$ is given in Eq.~\eqref{def:cov}. Now using the fact that $\Omega^T\Omega=\bbbone=\Omega\Omega^T$, we find from Eq.~\eqref{24.1} that 
\begin{equation}
\label{24.2}
\tfrac12 {\rm Var}_\sigma\big(\varphi(z)\big) = Z^T \Omega^T A \Omega Z,
\end{equation}
where $A$ is precisely the covariance matrix of $\sigma$, defined in Eq.~\eqref{eq:Adef}. Combining this with Eq.~\eqref{eq:lambda0limit} we  conclude that  the state $\rho_\infty^0$ is the   centered Gaussian state  having the same covariance matrix $A$ as the single-mode reservoir state $\sigma$.

 Suppose now the reservoir states $\sigma$ are stationary, so that their covariance matrix defined in Eq.~\eqref{eq:Adef} is anti-diagonal.  The corresponding Gaussian state is then a thermal state. This immediately leads to the following result.

\begin{prop}
\label{ATE}
When a system mode in an arbitrary initial state $\rho$ passes through a reservoir, the modes of which are all in the same stationary (but not necessarily thermal) state  $\sigma$, the asymptotic system state associated with the $a$-mode will, as the transmittance $\tau=\cos\lambda$ of the beam splitters governing the interaction increases towards unity, approach a thermal state having the same average photon number and energy as that of each of the reservoir modes through which it has passed.
\end{prop}

Unlike the result summarized in Proposition~\ref{RTE}, which demonstrates the return to equilibrium that the small system
undergoes as it passes through a thermal reservoir, 
we find that a repeated sequence of sufficiently weak interactions with the elements of a stationary but non-thermal reservoir suffices to drive the system mode to thermal equilibrium, at a temperature consistent with equipartition of the  energy of the entire system.
This is the main result of our analysis. 

A similar phenomenon of approach to equilibrium has been shown to occur in classical systems where a particle moves through an array of scatterers that are not in equilibrium and with which it can exchange energy and momentum. At strong coupling, the array will then drive the particle to a stationary state that will not, in general, be a state of thermal equilibrium. In the limit of small coupling, however, this asymptotic state will approach a thermally equilibrated state compatible with equipartition~\cite{DBMePa16}.

To further round out the analysis presented above, we note that when the state $\sigma$ is not centered, we can translate $\rho_\infty^\lambda$ in order to center it, i.e., we can define the centered state 
\begin{equation}\label{eq:centered}
\rho_{\infty, *}^\lambda
= D_a\left(-\frac{s}{1-c}\langle b\rangle_\sigma\right)\rho_\lambda^\infty D_a\left(\frac{s}{1-c}\langle b\rangle_\sigma\right),
\end{equation}
for which
$$
\chi_{\infty, *}^\lambda=\exp\left(-\frac{s}{1-c}\langle\varphi(z)\rangle_\sigma\right)\chi_\infty^\lambda(z).
$$
It then follows again from Eq.~\eqref{eq:lambdadevelop} that the asymptotic state
$\rho^0_{\infty, *}$ is the centered Gaussian state with the same covariance matrix as $\sigma$.

To end this section we briefly explain the link between our results, the   van Hove limit, and the quantum central limit theorem as discussed in~\cite{CuHu71, BeDaLa21}. For that purpose it is helpful to consider a slightly more general situation where the coupling parameter $\lambda$ is allowed to be different from one beam splitter to the next. Writing $S_k=\exp(-i\lambda_k(a^\dagger b_k+a b_k^\dagger))$, one has, in the Heisenberg picture,
\begin{eqnarray}\label{eq:QCLT}
a_K&:=&S^\dagger_KS^\dagger_{K-1}\dots S_2^\dagger S_1^\dagger a S_1 S_2\dots S_{K-1} S_K\nonumber\\
&=& (\Pi_{k=1}^K c_k)a+ i\sum_{k=1}^K s_k b_k,
\end{eqnarray}
where, as before, $c_k=\cos \lambda_k, s_k=\sin \lambda_k$. Since the first term tends to zero, one sees that, essentially, the annihilation operator $a_K$ of the  system mode of the beam after $K$ beam splitters, is a sum of the independent random variables given by the $s_kb_k$, since all modes of the reservoir are in the same initial state $\sigma$. If for fixed $K$ one now chooses $\lambda_k=1/\sqrt{K}$ for all $1\leq k\leq K$, then one is clearly in the situation of the central limit theorem as discussed in~\cite{CuHu71}. Note that this also corresponds precisely to the well known van Hove  limit, in which $K\to+\infty$ while $\lambda^2 K$ remains constant. It is easy to see that the proof of Theorem~\ref{thm:gaussianlim} of Appendix~\ref{app:approach} simplifies in this case and that, provided $\sigma$ has vanishing first moments,
$$
\lim_{\lambda\to0} \rho^\lambda_K=\rho_G,
$$
where $\rho_{\textrm G}$ is the Gaussian state with the same variance as $\sigma$.
Alternatively, one can take
$$
\lambda_k=\frac{1}{\sqrt{k+1}}
$$
for all $k$, independently of $K$. The limit $K\to+\infty$ can in this case be taken in the same manner as in the previous section, and the asymptotic state is again a Gaussian. This is the situation studied in~\cite{BeDaLa21}, where the rate of convergence to the asymptotic state is analyzed. 
Note that in these approaches, the limit $\lambda\to0$ and $K\to\infty$ are taken simultaneously, whereas we consider here the more natural regime where first $\lambda$ is kept fixed while $K$ is taken to infinity, leading to the asymptotic state $\rho_\infty^\lambda$, which is not necessarily Gaussian. Then only is $\lambda$ taken to be small.


\section{Entanglement and (non)classicality of the asymptotic state}\label{s:entnonclas}
In this section we study two typically quantum mechanical features of the asymptotic state $\rho_\lambda^\infty$. Specifically, we evaluate the degree to which the $a$-mode and the reservoir modes are asymptotically entangled, as well as the degree to which the state $\rho_\lambda^\infty$ of the $a$-mode is nonclassical.

We focus on the case in which all system and reservoir modes are initially in pure states, so that the entire system state is also initially pure. Since the evolution is unitary, this purity of the entire system state is preserved, and is thus also a feature of the entire system state after all   the scattering processes have occurred. 

Under these circumstances, the purity $\mathcal P^\lambda_\infty =
\Tr(\rho^\lambda_\infty)^2$ of the asymptotic state $\rho^\lambda_\infty$ provides a faithful measure of the asymptotic entanglement of the $a$ mode with the reservoir. It is easily computed to leading order in $\lambda$ by remarking first that the purity of  $\rho_\infty^\lambda$ is the same as the purity of the centered state $\rho_{\infty, *}^\lambda$, which, as we have seen, converges to a centered Gaussian as $\lambda$ goes to zero. Hence
\begin{eqnarray}
\label{eq:purityasymp}
\mathcal P^0_\infty&=&\lim_{\lambda\to0}\Tr (\rho^{\lambda}_{\infty })^2=\lim_{\lambda\to0}\Tr(\rho_{\infty, *}^{\lambda })^2\nonumber\\
&=&\Tr\rho_{\textrm G}^2=\det( V_\sigma)^{-1/2},
\end{eqnarray}
where $\rho_G$ is the centered Gaussian state with quadrature covariance matrix $V_\sigma$, given by
$$
V_\sigma=2
\begin{pmatrix}
\covsigma{X,X}&\covsigma{X,P}\\
\covsigma{P,X}&\covsigma{P,P}
\end{pmatrix}.
$$
Here the quadratures $X,P$ are defined as $X=\frac1{\sqrt2}(a^\dagger+a), P=\frac{i}{\sqrt 2}(a^\dagger-a)$ and the last equality in Eq.~\eqref{eq:purityasymp} is a known property of Gaussian states (see, for example \cite{Serafini}).
From the Schr\"odinger-Robertson uncertainty relation, which asserts that $\det V_\sigma\geq 1$ for all $\sigma$, and the fact that only Gaussian pure states saturate this inequality~\cite{Serafini}, one concludes that the asymptotic state  of the $a$ mode is entangled with the reservoir if and only if $\sigma$, which we recall is supposed pure,  is not Gaussian. 
In that situation, the von Neumann entropy 
$$
S(\rho_\infty^\lambda)=-\Tr (\rhoinfl\ln\rhoinfl)
$$ 
of the asymptotic state $\rho^\lambda_\infty$ is a measure of the entanglement of the $a$-mode and the reservoir. It can be similarly  evaluated to leading order in $\lambda\to0$: 
\begin{equation}\label{eq:vNasymp}
S_\infty^0:=\lim_{\lambda\to 0} S(\rhoinfl)=-\Tr(\rho_{\textrm G}\ln \rho_{\textrm G})=g(\sqrt{\det V_\sigma}),
\end{equation}
with~\cite{Serafini}
$$
g(x)=\frac{x+1}{2}\ln(\frac{x+1}{2})-\frac{x-1}{2}\ln(\frac{x-1}{2}).
$$
Gaussian states are known to maximize the von Neumann entropy among all states with a given covariance matrix~\cite{Wolf_2006}. This means that when both the initial system state $\rho$ and the initial reservoir mode state $\sigma$ are pure, and the coupling is small,  the repeated scattering process drives the system mode to the state with maximal entanglement with the reservoir under the constraint that its  covariance matrix equals that of the reservoir modes.

We have already remarked that, when the reservoir states $\sigma$ are stationary, the small coupling asymptotic state $\rho_\infty^0$ is thermal. In that case, this asymptotic state is therefore classical, in the precise sense that it is a convex mixture of coherent states~\cite{titulaer_correlation_1965}. For general $\sigma$, however, the asymptotic state is Gaussian and not necessarily classical in this sense. We now investigate how strongly nonclassical the asymptotic state can be. As above, we concentrate on the case where $\sigma=|\psi\rangle\langle\psi|$ is pure. There exists a large variety of nonclassicality measures and witnesses, among which Wigner negativity~\cite{kenfack_negativity_2004} is a popular choice. However, since the asymptotic state is Gaussian, it is Wigner positive. This means that the reservoir does not transfer or imprint any of its potential Wigner negativity on the $a$-mode in the repeated scattering process. It could however still transfer other nonclassical features. To evaluate this phenomenon, another nonclassicality measure is therefore needed.  We choose to use the quadrature coherence scale (QCS), introduced in~\cite{de_bievre_measuring_2019, HeDeB19}, and which has shown its efficiency as a nonclassicality measure on large families of benchmark states~\cite{Hoetal19, hertz_decoherence_2023}. It has also been shown to be experimentally measurable~\cite{goldberg_measuring_2023} using a protocol proposed in~\cite{griffet2022interferometric}. The QCS of a single mode state $\rho$ is defined as
$$
\Ccal^2(\rho)=\frac1{2\mathcal{P}}\left(\tr[\rho, X][X,\rho]+ \tr[\rho, P][P,\rho]\right),
$$
where $\mathcal P=\tr \rho^2$ is the purity of $\rho$.
 As its name indicates, the QCS is a measure of the scale on which the coherences $\rho(x,x')=\langle x|\rho|x'\rangle$ and $\rho(p,p')=\langle p|\rho|p'\rangle$ of the state $\rho$ are sizeable~\cite{HeDeB19}. 
The QCS is a nonclassicality witness since, when $\rho$ is nonclassical, $\Ccal^2(\rho)\geq 1$. In addition,  a large value of $\Ccal^2(\rho)$ is an indication of strong nonclassicality of the state $\rho$~\cite{de_bievre_measuring_2019}. Note that the QCS is translationally invariant. For a Gaussian state $\rho_{\textrm G}$, one finds~\cite{hertz_relating_2020} 
$$
\Ccal^2(\rho_{\textrm G})=\frac12 \Tr V^{-1}_{\rho_{\textrm G}}.
$$
For an arbitrary pure state $\sigma=|\psi\rangle\langle \psi|$ (Gaussian or not), one has
$$
\Ccal^2(\sigma)=\Delta X^2+\Delta P^2=\frac12\Tr V_\sigma.
$$
Since, for a pure state $\sigma=|\psi\rangle\langle \psi|$, 
$$
\Delta X^2+\Delta P^2=\frac12\det V_{\sigma} (\Tr V_{\sigma}^{-1}),
$$
it follows from the uncertainty relation in the form $\det V_{\sigma}\geq 1$ that 
$$
\Ccal^2(\sigma)\geq \Ccal^2(\rho_{\textrm G}),
$$
where $\rho_{\textrm G}$ is the Gaussian state with covariance matrix $V_{\sigma}$. 

From these general considerations it follows that
\begin{equation}\label{eq:QCSasymp}
\Ccal^{2, 0}_\infty:=\lim_{\lambda\to0}\Ccal^2(\rho_\infty^\lambda)=\Ccal^2(\rho_{\textrm G})=\frac{\Ccal^2(\sigma)} {\det V_\sigma}\leq \Ccal^2(\sigma).
\end{equation}
Eq.~\eqref{eq:QCSasymp} shows that  the scattering process imprints a fraction only of the QCS, hence of the nonclassicality, of the reservoir states on the asymptotic state. This fraction will be small if $\det V_\sigma$ is large. In that case, the purity of the asymptotic state is also small, see Eq.~\eqref{eq:purityasymp}, and its von Neumann entanglement entropy is large, see Eq.~\eqref{eq:vNasymp}. In other words, the entanglement of the $a$-mode with the reservoir is large, while its nonclassicality is small. One has in fact
$$
\Ccal^{2,0}_\infty= (\mathcal P_\infty^0)^2
\Ccal^2(\sigma).
$$
More precisely, one may note that Eq.~\eqref{eq:vNasymp} implies that the von Neumann entropy of the asymptotic state is a slowly growing function of $\det V_\sigma$. 
Indeed, for large $x$, we have
$$
g(x)\simeq \ln(\frac{x}{2}) +1, \quad x\simeq \frac{2}{\textrm e}\exp(g).
$$
So
\begin{eqnarray*}
\Ccal^{2,0}_\infty&=&\lim_{\lambda\to0}\Ccal^2(\rho_\infty^\lambda)=\Ccal^2(\rho_{\textrm G})\\
&\simeq& \left(\frac{\mathrm e}{2}\right)^2\Ccal^2(\sigma) \exp(-2S_\infty^0),
\end{eqnarray*}
and
$$
\mathcal P^0_\infty=\det( V_\sigma)^{-1/2}\simeq \frac{\mathrm e}{2}\exp(-S_\infty^0).
$$
As an example, when $\sigma=|n\rangle\langle n|$, one finds
$$
\Ccal^2(\sigma)=2n+1,\quad\Ccal^{0,2}_\infty=\frac1{2n+1},
$$
$$ \mathcal P_\infty^0=\frac1{2n+1},\quad S_\infty^0=g(2n+1).
$$
In other words, the more nonclassical $\sigma$ is (large $n$), as measured by the QCS, the more classical $\rho_\infty^0$ is. The entanglement of the $a$-mode with the reservoir, on the other hand, grows slowly (logarithmically) with $n$.

\bigskip

\noindent {\it Acknowledgments}: 
This work was supported in part by the Agence Nationale de la Recherche under grant ANR-11-LABX-0007-01 (Labex CEMPI), by the Nord-Pas de Calais Regional Council and the European Regional Development Fund through the Contrat de Projets \'Etat-R\'egion (CPER), and by  the CNRS through the MITI interdisciplinary programs. SDB thanks J.C. Garreau and N. Cerf for useful discussions. MM was supported by a Discovery Grant from NSERC (Natural Sciences and Engineering Research Council of Canada) as well as a Simons-CRM fellowship (Centre de Recherches Math\'ematiques, Montreal).
PEP thanks V.M. Kenkre from whom, long ago and in a place far away, he first learned about this subject.

\appendix

\section{Proofs}

\subsection{Existence of the asymptotic state} 
\label{App:A}
It is the goal of this appendix to give a precise proof of Eq.~\eqref{rhoinfty} and Eq.~\eqref{11}. For that purpose, we need the regularity assumption (A1) on the characteristic function $\chi_\sigma(z)$ of the reservoir modes, stated below.  

Given a density matrix $\sigma$ and a fixed complex number $z\in\mathbf C$, we introduce the function 
\begin{equation*}
	\mathbf R\ni x\mapsto \chi_{\sigma}(xz) = \av{D_{b}(x z)}_\sigma\in \mathbf C.
\end{equation*}
The assumption then reads as follows:
\begin{itemize}
	\item[{\bf (A1)}]\label{condition:1} There is a $\delta>0$ such that for every $z\in\mathbf C$ and every $x$ with $|x|<\delta$,  
	\begin{itemize}
		\item[--] The function $x\mapsto \chi_{\sigma}(xz)$ is differentiable.
		\item[--] There is a (possibly $z$-dependent) constant  $C_0(z)$ such that 
		$$
		|\tfrac{d}{d x}\chi_{\sigma}(xz)|\le C_0(z).
		$$ 
		\item[--] There is an $\eta>0$ such that for all $|z|<\eta$, we have $C_0(z)\le C_0$. 
	\end{itemize}
\end{itemize}
This technical condition is typically satisfied for many states considered. This includes Gaussian states, Fock states, and cat states. 

\begin{thm} \label{thm:existence}
Suppose that the condition {\rm (A1)} holds. Then, for all $0< \lambda <\pi/2$,  there exists a density matrix $\rho_\infty^\lambda$ so that for all $z\in\mathbf C$,
\begin{eqnarray}
\label{eq:rhoinftylambda}
\lefteqn{\Tr(\rho_\infty^\lambda  D_{a}(z)) =}\nonumber\\ &&\lim_{K\rightarrow\infty}\  \prod_{k=0}^{K-1} \av{D_{b}(\sin(\lambda)\{\cos(\lambda)\}^k z)}_\sigma.
	\end{eqnarray}
\end{thm}

\medskip

\noindent
{\it Proof.}  We first show that the limit $K\rightarrow\infty$ in Eq.~\eqref{11} exists. Recall the abbreviation $s=\sin\lambda$, $c=\cos\lambda$. Let $z\in \mathbf C$ be fixed. As $c<1$ there is an integer $k_0$ such that $c^k<\delta$ for all $k\ge k_0$. We use the fundamental theorem of calculus for the function $x\mapsto \ln[\chi_{\sigma}(sx z)]$, to get, for $k\ge k_0$,
\begin{eqnarray}
	\label{16}
	\ln [\chi_{\sigma}(sc^k z)] &=& \int_0^{c^k}  \frac{d}{dy}\ln[ \chi_{\sigma}({ s yz})] dy \nonumber\\
 &=&\int_0^{c^k} \frac{\frac{d}{dy} \chi_{\sigma}( s yz)}{\chi_{\sigma}( syz)} dy.
\end{eqnarray}
Since $\chi_{\sigma}(0)=1$ another application of the fundamental theorem of calculus gives
\begin{equation}
	\label{17}
	\chi_{\sigma}(syz)-1 = \int_0^{y} \frac{d}{dw}\chi_{\sigma}( swz)dw.
\end{equation}
Due to $|\frac{d}{dw}\chi_{\sigma}(swz)|\le C_0(sz)$ we obtain from Eq.~\eqref{17} that $|\chi_{\sigma}(syz)-1|\le y C_0( sz)\le c^kC_0( sz)$. Hence there is a $k_1$ (depending on $sz$) such that for $k\ge k_1$, we have
\begin{equation}
	\label{18}
	|\chi_{\sigma}(syz)|\ge 1/2
\end{equation}
for all $0\le y\le c^k$. Using Eq.~\eqref{18} we obtain from Eq.~\eqref{16} the bound,
\begin{equation}
	\label{19}
	\big| \ln[\chi_{\sigma}(s c^k z)]\big| \le 2c^kC_0(sz),\qquad k\ge k_1.
\end{equation}
This shows that the series $\sum_{k\ge 0} \ln [\chi_{\sigma}( sc^k z)]$ converges absolutely, which implies that the limit of the infinite product in Eq.~\eqref{11} exists.  

Next we show that the series converges {\it uniformly} in $z$ for $|z|\le \eta$. Once we know this we conclude that $z\mapsto \sum_{k\ge 0}\ln[\chi_{b,\sigma}( s c^kz)]$ is a continuous function of $z$ for $|z|\le\eta$, so that $\prod_{k\ge 0}\chi_{b,\sigma}( s c^kz) = \exp\{\sum_{k\ge 0}\ln[\chi_{b,\sigma}( s c^kz)]\}$ is also continuous in this domain. Let us address the uniform convergence now. The relations Eq.~\eqref{16}, Eq.~\eqref{17} are still valid for $k\ge k_0$ (with $k_0$  independent of $z$). For $|z|<\eta$ we have $|s z|\le \eta$ and so Eq.~\eqref{17} implies $|\chi_{b,\sigma}( s wz)-1|\le c^k C_0$, where the right hand side is now independent of $z$ for $|z|<\eta$. Then the bound Eq.~\eqref{18} is valid for all $k\ge k_2$ with a $k_2$ independent of $z$ and so we get, analogous to Eq.~\eqref{19}, $| \ln[\chi_{b,\sigma}( s c^k z)] | \le 2c^kC_0$, $k\ge k_2$, uniformly in $|z|<\eta$. Now $\sum_{k\ge 0} \ln[\chi_{b,\sigma}(s c^k z)]$ converges absolutely and uniformly in $|z|<\eta$ by the Weierstrass $M$-test.

We have shown so far that the limit as $K\rightarrow\infty$ in Eq.~\eqref{10} exists. This means that the limit of the characteristic function associated to the density matrix $L^K(\rho)$ exists as $K\rightarrow\infty$. Moreover, we have shown that this limit characteristic function, $\chi_\infty^\lambda$, is continuous in $z$ at the origin $z=0$. It is known \cite{Lami2018} that then, $\chi_\infty^\lambda$ corresponds to a limit density matrix $\rho_\infty^\lambda$, meaning that there is a density matrix $\rho_\infty^\lambda$ such that $\chi_\infty^\lambda(z)=\Tr (\rho_\infty^\lambda D_a(z))$, and that furthermore, $L^K(\rho) \rightarrow \rho_\infty^\lambda$ in trace norm, as $K\rightarrow\infty$. \hfill  \qedsymbol{}

\subsection{Approach to equilibrium}
\label{app:approach}

The main result of this section is Theorem \ref{thm:gaussianlim}, which we have  used  in Section \ref{s:smallcoupling}. With the definition Eq.~\eqref{varphidef} of $\varphi(z)$ we have the Taylor series expansion,
\begin{equation*}
	D_b(z) = e^{\varphi(z)} = \bbbone +\varphi(z) +\tfrac12\varphi(z)^2+\cdots 
\end{equation*}
We now impose a regularity condition on the initial single-mode reservoir state $\sigma$:
\begin{itemize}
	\item[{\bf (A2)}] Suppose $\av{b}_\sigma$, $\av{b^\dag b}_\sigma$ and $\av{b^2}_\sigma$ are finite. Moreover, suppose there is a $c_0>0$ such that for all $|z|<c_0$,
	\begin{equation}
		\label{26.1}
		\av{D_b(z)}_\sigma = 1 +  \av{\varphi(z)}_\sigma +\tfrac{1}{2}\av{\varphi(z)^2}_\sigma +R_\sigma(z),
	\end{equation}
	where $|R_\sigma(z)| \le C |z|^3$ for some constant $C$.
\end{itemize}
Recall the assumption (A1), given before Theorem \ref{thm:existence}. Our main result of this section is:

\noindent
\begin{thm}\label{thm:gaussianlim}
Suppose $\sigma$ satisfies {\rm (A1)} and {\rm (A2)} and has vanishing first moment, $\av{b}_\sigma=0$. Then the limit 	
$$
\lim_{\lambda\rightarrow 0}  \rho_\infty^\lambda = \rho_\infty^0
$$
exists and $\rho_\infty^0$ is the centered Gaussian state which has the same covariance matrix as the state $\sigma$. 
	
Moreover, if the moment $\av{b}_\sigma$ does not vanish, then $\rho_\infty^\lambda$ does not have a limit as $\lambda\rightarrow 0$. 
\end{thm}

\noindent{\it Proof of Theorem \ref{thm:gaussianlim}.\ } 
Condition (A2) means that  $\av{\varphi(z)}_\sigma$ and $\av{\varphi(z)^2}_{\sigma}$ are finite and
\begin{equation}
	\label{26}
	\av{e^{\varphi(z)}}_\sigma = 1 +\av{\varphi(z)}_\sigma+\tfrac12 \av{\varphi(z)^2}_\sigma +R_\sigma(z).
\end{equation}
The proof of Theorem \ref{thm:gaussianlim} is based on the following result.

\begin{prop}
	\label{thm2}
	Suppose $\sigma$ satisfies the assumptions {\rm (A1)} and {\rm (A2)}. For each $z\in\mathbf C$ there is a $\lambda_0>0$ such that if\,  $0<\lambda<\lambda_0$, then  
	\begin{eqnarray}
		\label{27.1}
\lefteqn{\tr\big(\rho_\infty^\lambda D_a(z)\big)}\\
&=& \exp\Big[  2\lambda^{-1}\av{\varphi( z)}_\sigma +\tfrac12 {\rm Var}_\sigma\big(\varphi( z)\big) +\lambda t(\lambda,z)\Big],
\nonumber
	\end{eqnarray}
	where ${\rm Var}_\sigma(X)=\av{X^2}_\sigma-\av{X}_\sigma ^2$ and where the remainder term $t(\lambda, z)$ satisfies $|t(\lambda,z)|\le C(z)$ for a constant $C(z)$. 
\end{prop}

We give a proof of Proposition \ref{thm2} below. For now we use the result to show Theorem \ref{thm:gaussianlim}. First, if $\av{\varphi(z)}_\sigma\neq 0$, then Eq.~\eqref{27.1} shows that the average of $D_a(z)$ in $\rho_\infty^\lambda$ does not have a limit as $\lambda\rightarrow 0$. This means that $\rho^\lambda_\infty$ does not have a limit. Next suppose $\av{\varphi(z)}_\sigma=0$ for all $z$. Then according to Eq.~\eqref{27.1}
\begin{equation}
	\label{28}
	\lim_{\lambda\rightarrow 0} \tr\big(\rho_\infty^\lambda D_a(z)\big) = \exp\Big[  \tfrac12 {\rm Var}_\sigma\big(\varphi(z)\big) \Big].
\end{equation}
By condition (A2), the map $z\mapsto {\rm Var}_\sigma(\varphi(z))$ is continuous at the origin. Eq.~\eqref{28} means that the characteristic function of $\rho_\infty^\lambda$ has a limit as $\lambda\rightarrow 0$, and this limit is continuous in $z$ at the origin $z=0$. It follows from \cite{Lami2018} (``SWOT convergence Lemma'') that $\rho_\infty^\lambda$ converges in trace norm to some density matrix we denote $\rho_\infty^0$, as $\lambda\rightarrow 0$, and that moreover, the characteristic function of $\rho_\infty^0$ is the limit characteristic function of the $\rho_\infty^\lambda$. In other words, for all $z\in\mathbf C$, 
\begin{eqnarray*}
\Tr\big(\rho_\infty^0 D_a(z)\big) &=&\exp\big[  \tfrac12 {\rm Var}_\sigma\big(\varphi(z)\big) \big].
\end{eqnarray*}
This shows that $\rho_\infty^0$ is Gaussian. By proceeding as in Eq.~\eqref{24.1}-\eqref{24.2}, we identify $\rho_\infty^0$ as the centered Gaussian having the same covariance matrix $A$ as $\sigma$. 
This completes the the proof of Theorem \ref{thm:gaussianlim}, modulo a proof of Proposition \ref{thm2}, which we give now.

\noindent{\em Proof of Proposition \ref{thm2}.\ } Theorem \ref{thm:existence} gives the limit state expectation functional as 
\begin{eqnarray*}
	\tr\big(\rho_\infty^\lambda D_a(z)\big) &=& \prod_{k\ge0} \av{\e^{\varphi(sc^kz)}}_\sigma \\
 &=& \exp\big[\sum_{k\ge 0}\ln\av{\e^{ \varphi( \epsilon_k z)}}_\sigma \big],
\end{eqnarray*}
where  employed $D_b(z)=\e^{\varphi(z)}$ and we set, for notational simplicity,
\begin{equation}
	\label{zeta}
	\epsilon_k =sc^k,\qquad s=\sin(\lambda),\quad c=\cos(\lambda).
\end{equation}
Choose $\lambda$ small enough (depending on $z$) such that $s|z|<c_0$, where $c_0$ is the constant appearing in Assumption (A2).  We expand using Eq.~\eqref{26},
\begin{eqnarray}
	\label{29}
	\ln\av{\e^{\varphi( \epsilon_k z)}}_\sigma &=&  \ln\big[ 1 + \epsilon_k \av{\varphi( z)}_\sigma +\tfrac{\epsilon_k^2}{2}\av{\varphi(z)^2}_\sigma \nonumber\\
 &&\quad +R_\sigma(\epsilon_k z)\big] \nonumber\\
	&=& \sum_{n\ge 1} (-1)^{n+1}\frac{\xi_k^n}{n},
\end{eqnarray}
where we set
\begin{equation}
	\label{xi}
	\xi_k = \epsilon_k \av{\varphi(z)}_\sigma +\tfrac{\epsilon_k^2}{2}\av{\varphi(z)^2}_\sigma +R_\sigma(\epsilon_k z).
\end{equation}
The power series for the logarithm in Eq.~\eqref{29} converges (absolutely) since 
\begin{equation}
	\label{smallxi}
	|\xi_k| \le s c^kC(z)\le s C( z) <1
\end{equation}
for $s$ small enough (that is, $\lambda$ small enough, with an upper bound possibly depending on $z$), and where $C(z)$ is some constant not depending on $k$. We split off the main terms in the series Eq.~\eqref{29},
\begin{equation}
	\label{30-1}
	\sum_{n\ge 1} (-1)^{n+1}\frac{\xi_k^n}{n} = \xi_k -\tfrac12 \xi_k^2 +\sum_{n\ge 3}(-1)^{n+1}\frac{\xi_k^n}{n}
\end{equation}
and we estimate the infinite sum with $n\ge 3$ as
\begin{eqnarray}
	\label{32}
	\Big|\sum_{n\ge 3}(-1)^{n+1}\frac{\xi_k^n}{n}\Big|&\le& \sum_{n\ge 3}|\xi_k|^n=\frac{|\xi_k|^3}{1-|\xi_k|}\nonumber\\
 &\le& s^3 c^{3k}C(z),
\end{eqnarray}
provided that $s$ is small enough (with an upper bound possibly depending on $z$), and where $C(z)$ is a constant independent of $k$. By using Eq.~\eqref{xi} the linear and quadratic terms in Eq.~\eqref{30-1} satisfy the bound 
\begin{multline}
	\label{34}
\Big| \xi_k -\tfrac12\xi_k^2 - \big\{\epsilon_k \av{\varphi(z)}_\sigma +\tfrac12 \epsilon_k^2 {\rm Var}_\sigma(\varphi(z))  \big\}\Big|\\
\le s^3 c^{3k}C(z)
\end{multline}
for a constant $C(z)$ not depending on $k$.  Combining Eq.~\eqref{29}, Eq.~\eqref{30-1}, Eq.~\eqref{32}, and Eq.~\eqref{34} gives
\begin{multline}
	\label{36}
	\Big| \ln\av{\e^{\varphi(\epsilon_k z)}}_\sigma -\big\{\epsilon_k \av{\varphi(z)}_\sigma +\tfrac12 \epsilon_k^2 {\rm Var}_\sigma(\varphi(z))  \big\}\Big|\\
 \le s^3 c^{3k}C(z)
\end{multline}
provided  $s$ is small enough (with an upper bound possibly depending on $z$), and where $C(z)$ is a constant independent of $k$. The bound Eq.~\eqref{36} shows that there exists a $\lambda_0$ (possibly depending on $z$) such that whenever $\lambda\le \lambda_0$, then we have, for any integer $K$, 
\begin{multline}
	\label{37}
	\Big| \sum_{k=0}^K \Big[ \ln\av{\e^{ \varphi(sc^k z)}}_\sigma -\big\{ sc^k \av{\varphi(z)}_\sigma \\+\tfrac12 s^2c^{2k} {\rm Var}_\sigma(\varphi(z))  \big\}\Big]\Big| 
	\le s^3  C(z)\sum_{k=0}^K c^{3k}. 
\end{multline}
By taking $K\rightarrow\infty$ we get
\begin{multline*}
	\sum_{k\ge 0} \ln\av{\e^{ \varphi(sc^k z)}}_\sigma =\\
 \frac{s}{1-c}\av{\varphi(z)}_\sigma +\tfrac12 {\rm Var}_\sigma(\varphi(z)) +t(s,z)  
\end{multline*}
with $|t(s,z)| \le C(z)\frac{s^3}{1-c^3}\le C(z) s\le C(z)\lambda$ (where we use the symbol $C(z)$ for a constant which can vary from bound to bound). As $\frac{s}{1-c}=2/\lambda +O(\lambda)$ for small $\lambda$, this completes the proof of Proposition \ref{thm2} and hence this completes the proof of Theorem \ref{thm:gaussianlim}. \hfill $\square$



\end{document}